# benchNGS : An approach to benchmark short reads alignment tools

Farzana Rahman[a], Mehedi Hassan[a], Alona Kryshchenko[b], Inna Dubchak[d], Tatiana V. Tatarinova[b], Nickolai Alexandrov[c]

[a] *University of South Wales, Treforest, UK* [b] *University of Southern California, Los Angeles, USA* [c] *International Rice Research Institute, Los Banos, Philippines*
[d] *US Department of Energy Joint Genome Institute (JGI), Walnut Creek, USA*

Abstract

In the last decade a number of algorithms and associated software have been developed to align next generation sequencing (NGS) reads with relevant reference genomes. The accuracy of these programs may vary significantly, especially when the NGS reads are quite different from the available reference genome. We propose a benchmark to assess accuracy of short reads mapping based on the pre-computed global alignment of related genome sequences.

In this paper we propose a benchmark to assess accuracy of the short reads mapping based on the pre-computed global alignment of closely related genome sequences. We outline the method and also present a short report of an experiment performed on five popular alignment tools based on the pairwise alignments of *Escherichia coli* O6 CFT073 genome with genomes of seven other bacteria.

*Keywords:* Benchmark, NGS, alignment, short reads, BLAST, SOAP, Bowtie, bwa, SHRiMP

Introduction

Next Generation Sequencing (NGS) technologies provide fast and cost-effective alternatives to the established Sanger sequencing, and powers impressive scientific achievements and development of novel biological applications in medicine, ecology, forensics, epidemiology and other fields of science [26, 27]. High throughput NGS technology comes with challenges in managing large datasets and the "big data" questions in biology. Open access publications and public domain data liberation, made way for development of a plethora of tools for analysis of these datasets. With hourly paid cloud-based computing services being increasingly available, researchers are now in need of a benchmark method to select the perfect tool, that is fit for purpose. Our endeavor is to establish a benchmark method for short read aligning tools.

De novo assembly of long sequence reads from Sanger-based sequencing process produces reliable genomic sequences [24]. Sanger sequence reads are typically 650 to 850 bases long while the NGS methods produce much shorter reads that are 50-450 bases long. The reads are assembled to chromosomes using well established algorithms, such as Celera Assembler[20], Arachne [2], Atlas, CAP3 [14], Euler [23], PCAP[15], Phrap [10, 11], RePS [30], Phusion [19]. Most of the assemblers follow the "overlap-layout-consensus" algorithmic strategy [22] or are based on a de Bruijn graph[6]. Usually, the "overlap" portion of the assembly process is the most computationally intensive. Using NGS reads for assembling whole genomes significantly reduces the costs of genome sequencing.

However, most of the existing sequence assembly programs are not suitable for short sequence reads generated by NGS methods [21]. This is partly because, the information contained in a short read is not sufficient to find a position of a read in a genome [32]. Moreover, the number of NGS reads is several orders of magnitude larger than the Sanger sequencing reads. For a novel or little-explored genomes, this can prove very difficult. Therefore, different algorithmic strategies more suitable for the short reads assembly have been developed. Usage of two sets of restriction enzymes creates overlapping libraries and reduces errors. It is also possible to use long and short reads together to take advantage of the low cost of NGS sequencing and computational unambiguity of long reads [31, 28, 7]. Finally, there is an "alignment-layout-consensus" approach that uses a reference genome. One of the implementations of this strategy is AMOS Comparative Assembler [24].

When a reference genome is used to guide a sequence assembly, the quality of the resulting assembly depends on the specific algorithm used, on frequency of repeats in the pair of genomes, and evolutionary distance between them. In addition, insertions in the target genome cannot be assembled using the "alignment-layout-consensus" approach and presence of rearrangements will negatively affect the quality of assembled contigs [24]. It has been demonstrated [24] that the "alignment-layout-consensus" approach works well for a pair of strains of the same bacteria (92-94% coverage of the target genome), but fails for more distinct sequences (11.4% coverage of the target genome using more diverse organisms such as *Streptococcus agalactiae* vs. *Streptococcus pyogenes*). *S. pyogenes* is a human pathogen, exclusively adapted to the human host, and *S. agalactiae* is one of the principal causes of bovine *Streptococcal mastitis* [18]. An array of computationally efficient tools for mapping of short reads onto reference genomes, such as SOAP, Bowtie, SHRiMP and BWA, has been developed. Well-established sequence alignment tools like BLAST [1] can also handle short reads alignment.

It is important to determine the limits of applicability of the reference-based alignment method depending on the divergence between the reference and target species. In this paper we chose a simulation approach using global whole genome alignments as gold standards. Simulation enables us to generate "NGS reads" of arbitrary length without investing in sequencing, map them to a reference genome and assess the correctness of a mapped position. To estimate error rate of these programs we propose a benchmark, which uses the large-scale alignment between syntenic regions of genome sequences as the true alignment. The aligned fragments of the whole genome alignment were cut into short sequence 'reads and the ability of different programs to reproduce the true alignment using these reads was tested. This proposed benchmark is a convenient way to select programs that are most suitable for the reference-based genome assembly. It gives clear, realistic and robust estimates of the accuracy of the alignment programs. The benchmark also defines the limits of sequence similarities for selecting a reference genome.

In this paper, we compare performance of the five popular freely available alignment programs using whole-genome alignment between between several strains of model bacteria *Escherichia coli* and between strains of *E. coli* and several species of *Salmonella*. We focused our analysis on bacterial species for a number of reasons. They have manageable size genomes, variety of nucleotide composition, and alignment of bacterial reads to genomes is essential for environmental and clinical applications, annotation of variants, determination of toxicity, drug resistance and pathogenicity of the analyzed strain [3],[29].

Proposed Methods

We propose the following procedures to institute the benchmark method. To evaluate the effectiveness of an aligner, we propose to compare the alignments done by the tool with a gold standard alignment from other independent sources. Researchers at various laboratories have invested significant efforts in obtaining a consensus global alignment among several model species. We intend to make use of these alignments to achieve a single benchmark score for a given tool.

Our procedure starts by extracting the reference genome and query genome from a peer reviewed global alignment. We call this the "Gold Standard Alignment (GSA)". We split the query sequence into short reads of pre-defined lengths. The rationale is that a "perfect" tool *per se*, will be able to align these small sequence fragments to their accurate alignment positions within the reference genome, replicating the results of GSA. The precision rate close to one will present a "near perfect" tool [25].

Different alignment tools present alignments in different formats. Our procedure does not rank the tools based on the developer's claim of accuracy. For example, the E-value reported by the BLAST tool is not used in the result of our scoring. We collect an information set, R from the alignment results containing: (i) read id ($r(n)$), (ii) reference sequence identifier (ref), (iii) start position of the read (stp). This information is then compared with their counterparts from the GSA.

To evaluate the quality of mapping of reads to the reference genome, we used a scoring method formed of True Positives (TP), False Positive (FP), False Negative (FN). When a short read or fragment is mapped exactly to the same position on the reference genome as defined by GSA, we award one point towards TP. If a fragment is mapped to a different position than the one defined by the global alignment, a penalty is awarded to FP. However, if the candidate tool failed to align a fragment to the correct location as GSA, then a penalty point is awarded to FN. To conclude benchmark of each candidate tool, we use Rijsbergen's $F1$ score as a measure of test accuracy [25].

We used true positive rate ($r$) and positive predictive rate ($p$), to compute the $F1$ score. Sensitivity or True Positive rate, also called "Recall", is computed by dividing of the total number of correct results by the number of alignments that were expected: $r = TP/(TP + FN)$.

Positive Predictive Value ("Precision")is computed by dividing of the total number of correct alignments by total number of alignments detected by the tool: $p = TP/(TP + FP)$.

The $F1$ score is computed from Precision and Recall, and it ranges from zero to one:
$F1 = 2 \times ((p \times r)/(p + r))$.

---

**Algorithm 1:** Benchmarking of a list of NGS Short Reads Aligner

**Data:** GSA: Gold Standard Alignment between two sequences
    Model, M: Reference genome of GSA
    Query, Q: Query genome of GSA
    Tools, T: List of the candidate tools

Initialization;
Data Preparation: Simulate short reads, $q(n) \subset Q$ of variant bp lengths $n \subset \{50, 100, 150, 400\}$;
**foreach** $t \subset T$ **do**
    **foreach** $q(n)$ **do**
        Align the reads to the model genome;
        From new alignment results generate R ← $\{q(n), ref, stp\}$ :
        Compare R with GSA and produce a set $S \leftarrow \{TP, FP, FN\}$ where
        True Positive Rate, $r \leftarrow \frac{TP}{TP+FN}$;
        Positive Predictive Value, $\rho \leftarrow \frac{TP}{TP+FP}$;
        Rijsbergen's accuracy measurement score, $F_1 = 2 * \frac{\rho*r}{\rho+r}$
    **end**
**end**
**Result:** Benchmark Score, $F_1$

Table 1: List of paired strains and their whole genome alignment statistics

| Genome | Accession | Identity | Al. Length |
|---|---|---|---|
| *S. enterica* Typhi Ty2 | NC 004631.1 | 56.58 | 29480 |
| *S. enterica* Typhi CT18 | NC 003198.1 | 54.43 | 29159 |
| *S. typhimurium* LT2 | NC 003197.1 | 58.02 | 29025 |
| *S. enterica* Paratyphi-A SARB42 | NC 006511.1 | 52.91 | 32221 |
| *E. coli* O157:H7 EDL933 | NC 002655.2 | 76.38 | 34335 |
| *E. coli* K12 | NC 000913.2 | 79.48 | 38457 |
| *E. coli* Sakai O157:H7 | NC 002695.1 | 77.46 | 34316 |

Implementation

We designed the computational experiment using model species *Escherichia sp.* and *Salmonella sp.* For our test cases we used the pre-computed global alignments of the following pairs of bacterial strains done by the VISTA consortium of Lawrence Berkeley National Laboratory and Joint Genome Institute [12, 9].

We used seven pairs of alignments between *Escherichia coli* O6 CFT073 and seven other strains of *Escherichia* and *Salmonella*. Table 1 contains a list of paired strains together with whole genome alignment statistics. Average percent identity is calculated as the number of identical nucleotides divided by the alignment length. Average alignment length is computed from all fragments in the corresponding whole-genome alignment.

*GSA Selection Justification*

We chose VISTA global alignments the GSA, since this technique generates long continuous DNA fragments of orthologous genomic intervals. VISTA uses a combination of global and local alignment methods consisting of three steps; (a) obtaining a map of large blocks of conserved synteny between the two species by applying Shuffle-LAGAN glocal chaining algorithm [5] to local alignments by translated BLAT [16]; (b) using Supermap [8], the fully symmetric whole-genome extension to the Shuffle-LAGAN [4], and (c) applying Shuffle-LAGAN the second time on each syntenic block to obtain a more fine-grained map of small-scale rearrangements.

*Short Reads Simulation*

As proposed in the method, to maintain consistency we used *Escherichia coli* O6 CFT073 genome as a reference genome. We used the second genome from each pairings as the query genome. Using a simple R program, we simulated short reads of lengths of $n$ bp where $n$=50, 100, 150, 400 from the reads. Each nucleotide was used as a start point of a new read as long as they ended with a read of expected length ($n$ bp).

Table 2: List of alignment tools used

| Tool Name | Version Used |
|---|---|
| BLAST+: NCBI Basic Local Alignment Search Tool | 2.2.26 |
| Bowtie 2: Bowtie Short Read Aligner | 2.1.0 |
| SHRiMP: SHort Read Mapping Package | 2.2.3 |
| SOAP2: Short Oligonucleotide Analysis Package | 2.2.1 |
| BWA: Burrows-Wheeler Aligner | 0.7.0 |

*Selection of Candidate Tools*

There is a large number of alignment tools available in the public domain. We intend to use most (if not all) of the tools to produce a comprehensive benchmark database. However, for this case study we used only a subset of the most popular alignment tools. Table 2 presents a list of the tools and their versions that were used. To maintain consistency, we did not use the latest versions of all the tools and rather dependent on the stable releases of the tools from a contemporary release time.

All of the tools were used as-is and without modification. Default parameters were used and the user guides were consulted only to install and run examples as recommended by developers.

Results and Discussion

The aim of the experiment was to examine how the tools perform with reads of varying lengths. Very short reads of 50 base pairs and relatively longer reads 400 base pairs were of special interest. We used the evolutional tree as a biological reference to observe accuracy of the benchmark. We expect that, if the genomes are identical, all five candidate programs should provide close alignments with high precision, yielding in F1 scores close to 1. Likewise, the tools are expected to yield a lower F1 scores for alignments performed between more distant organisms.

Our experiment demonstrated limits of sequence similarity for different programs. As expected, for alignments between various strains of same species (*E. coli*), all programs performed reasonably well, with the exception of SOAP2. For shorter reads of 50bp and 100bp, all five candidate tools demonstrated good F1 scores. However, as the reads lengths started to increase, at 150bp and 400bp, SOAP and BWA did not stay in-par with BLAST+ and SHRiMP.

For closely related genomes, BLAST+'s performance matched its reputation, however, for distant genomes, the performance was rather poor. For alignments between *Salmonella ep.* and *E. coli*, for reasonably shorter reads (50-100bp), BLAST+ was outperformed by SHRiMP. As the read lengths increased, BLAST+ showed a recovering trend.

In our experiment, Bowtie started with a below-par accuracy score for short reads, and with the increase of reads lengths, the accuracy continued to decrease.

For more distant species, SHRiMP performs significantly better. In almost all cases, SOAP showed the worst performance. Poor performance of SOAP can be explained by the fact that mapping of the reads in SOAP is mismatch-dependent. In an earlier study [13] it was observed that the suboptimal hits reduce from 21% to 1%, when mismatch rate was changed from 2 to 6 mismatches invoking the different behavior of the tool, which partially depends on the mismatch. More recently, it was demonstrated that SOAP has a lower read mapping accuracy in meta-genome experiments and it shows significant differences in the coverage depth [17], which agrees with our demonstrated results.

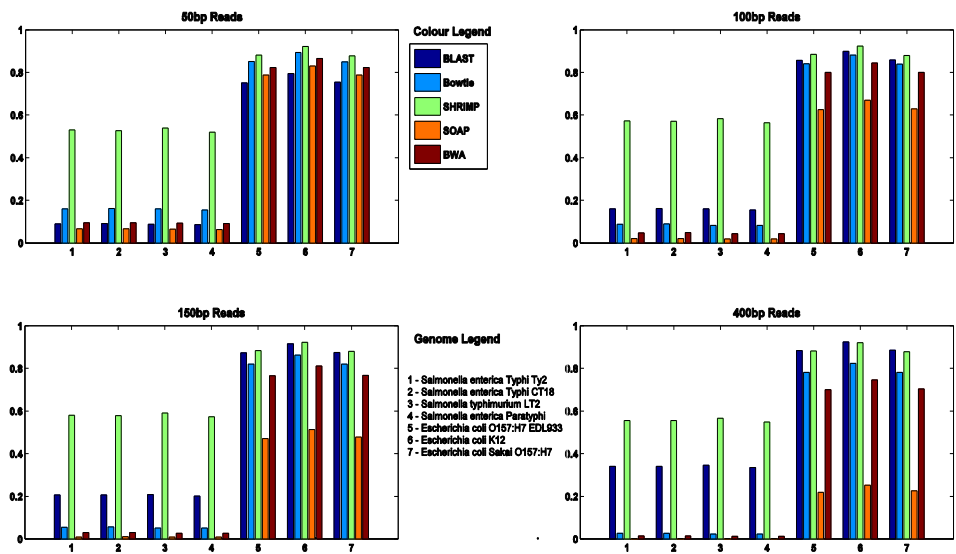

Figure 1: F1 Score for different read lengths.

CONCLUSIONS

We developed and experimented a benchmark strategy to assess the correctness of alignments produced by different tools. We tested our method on five tools and on a set of case study data. Our tested method proves our hypothesis about closely related genomes. If the genomes are identical, the tools perform well. If the genomes are distantly related by evolution such as E.coli and Salmonella, the tools perform differently. In our case, SHRiMP over-performs rest of the tools and SOAP performed reasonably bad. BLAST and Bowtie performed well after SHRiMP. BLAST showed consistent result as per our hypothesis. We conducted this experiment on a set of data to establish the benchmark method. We aim to extend our study for different species (i.e *Homo sapiens* vs *Pan troglodytes*) and adding a range of different tools for comparative analysis.

Availability of Supporting Data

The gold standard global alignments were collected from VISTA website available at : http://pipeline.lbl.gov/data/ecoli2/.

Simulated reads and outputs of BLAST, Bowtie, SHRiMP and SOAP are accessible via http://cbio.uk/benchNGS/. UNIX executable of a program created using this algorithm is also available at the same link.

ACKNOWLEDGEMENTS

The authors are grateful to Alexandre Poliakov for his work on whole-genome alignments.

**Funding**

FR was supported by HPC-Wales and Fujitsu Lab Europe. TT was supported by grants from The National Institute for General Medical Studies (GM068968), and the Eunice Kennedy Shriver National Institute of Child Health and Human Development (HD070996).

Authors Contributions

TT and AN conceptualized the benchmark and proposed initial framework. ID, FR, AK, MH designed the experiment, performed the case study, generated results and prepared manuscript. FR, TT, ID and NA interpreted the results and wrote the manuscript.